\documentclass[a4paper,12pt]{article}

\pdfoutput=1
\usepackage{amsmath}
\usepackage{amssymb}
\usepackage{mathrsfs}
\usepackage{bbm}
\usepackage{graphicx,subfigure}
\usepackage[numbers,sort&compress]{natbib}
\usepackage{array,longtable}
\usepackage{multirow}

\usepackage[utf8]{inputenc}

\usepackage[colorlinks,
            linkcolor=black,
            filecolor=black,
            anchorcolor=black,
            urlcolor=black,
            citecolor=black,
            bookmarks=false,
            ]{hyperref}

\numberwithin{equation}{section}

\def\gev{{\rm GeV}}

\newlength{\dinwidth}
\newlength{\dinmargin}
\allowdisplaybreaks

\begin{document}

\title{\bf \Large The $\boldsymbol{\Lambda_b\to \Lambda(\to p\pi^-)\mu^+\mu^-}$ decay in the aligned two-Higgs-doublet model}

\author{
Quan-Yi Hu\footnote{qyhu@mails.ccnu.edu.cn},
Xin-Qiang Li\footnote{xqli@mail.ccnu.edu.cn}\,
and
Ya-Dong Yang\footnote{yangyd@mail.ccnu.edu.cn}\\[15pt]
\small Institute of Particle Physics and Key Laboratory of Quark and Lepton Physics~(MOE), \\
\small Central China Normal University, Wuhan, Hubei 430079, China}

\date{}
\maketitle

\vspace{-0.2cm}

\begin{abstract}
{\noindent}The rare baryonic decay $\Lambda_b\to \Lambda(\to p\pi^-)\mu^+\mu^-$ provides valuable complementary information compared to the corresponding mesonic $b\to s\mu^+\mu^-$ transition. In this paper, using the latest high-precision lattice QCD calculation of the $\Lambda_b\to\Lambda$ transition form factors, we study this interesting decay within the aligned two-Higgs-doublet model, paying particularly attention to effects of the chirality-flipped operators generated by the charged scalars. In order to extract the full set of angular coefficients in this decay, we consider the following ten angular observables that can be derived from the analysis of the subsequent parity-violating $\Lambda\to p\pi^-$ decay: the differential branching fraction ${\rm d}{\cal B}/{\rm d}q^2$, the longitudinal polarization fraction $F_L$, the lepton-, hadron- and combined lepton-hadron-side forward-backward asymmetries $A_{\rm FB}^\ell$, $A_{\rm FB}^\Lambda$ and $A_{\rm FB}^{\ell\Lambda}$, as well as the other five asymmetry observables $Y_i$~($i={\rm 2,\,3s,\,3sc,\,4s,\,4sc}$). Detailed numerical comparisons are made between the SM and NP values for these angular observables. It is found that, under the constraints from the inclusive $B\to X_s\gamma$ branching fraction and the latest global fit results of $b\to s\ell\ell$ data, the contributions of right-handed semileptonic operators $O^{\prime}_{9,10}$, besides reconciling the $P_5^{\prime}$ anomaly observed in $B^0\to K^{\ast 0}\mu^+\mu^-$ decay, could also enhance the values of ${\rm d}{\cal B}/{\rm d}q^2$ and $A_{\rm FB}^\ell$ in the bin $[15,20]~{\rm GeV}^2$, leading to results consistent with the current LHCb measurements.
\end{abstract}

\newpage

\section{Introduction}
\label{sec:intro}

The rare semileptonic $b$-hadron decays induced by the flavour-changing neutral current (FCNC) transition $b\to s\ell^+\ell^-$ do not arise at tree level and, due to the Glashow-Iliopoulos-Maiani (GIM) mechanism~\cite{Glashow:1970gm}, are also highly suppressed at higher orders within the Standard Model (SM). In many extensions of the SM, on the other hand, new TeV-scale particles can participate in the SM loop diagrams and lead to measurable effects in these rare decays. As a consequence, they play an important role in testing the SM and probing New Physics (NP) beyond it~\cite{Hurth:2010tk,Blake:2016olu}.

While no any solid evidence of NP has been found in direct searches at high-energy colliders, it is interesting to note that several persistent deviations from the SM predictions have been observed in rare B-meson decays~\cite{Blake:2016olu}. Specific to the $b\to s\ell^+\ell^-$ mesonic decays, these include the angular observable $P_5'$ in the kinematical distribution of $B^0\to K^{\ast0}\mu^+\mu^-$~\cite{Aaij:2013qta,Aaij:2015oid,Abdesselam:2016llu,
DescotesGenon:2012zf,Descotes-Genon:2013vna}, the lepton-flavour-universality-violation ratio $R_K$ of the decay widths for $B\to K\mu^+\mu^-$ and $B\to K e^+ e^-$~\cite{Aaij:2014ora,Hiller:2003js,Bordone:2016gaq}, as well as the differential decay rates for $B\to K^{(\ast)}\mu^+\mu^-$~\cite{Aaij:2013iag,Aaij:2014pli,Bouchard:2013mia} and $B_s\to\phi\mu^+\mu^-$~\cite{Aaij:2013aln,Aaij:2015esa,Horgan:2013pva}. Motivated by these anomalies and using the other available data on such rare mesonic decays, several global analyses have been made~\cite{Descotes-Genon:2013wba,Altmannshofer:2013foa,Beaujean:2013soa,
Hurth:2013ssa,Altmannshofer:2014rta,Hurth:2014vma,Beaujean:2015gba,Du:2015tda,
Descotes-Genon:2015uva,Hurth:2016fbr}, finding that a negative shift in the Wilson coefficient $C_9$ improves the agreement with the data. However, due to the large hadronic uncertainties involved in exclusive modes, it remains quite unclear whether these anomalies indicate the smoking gun of NP, or are caused merely by underestimated hadronic power corrections~\cite{Hurth:2016fbr,Khodjamirian:2010vf,Khodjamirian:2012rm,Jager:2012uw,
Jager:2014rwa,Ciuchini:2015qxb,Lyon:2014hpa,Descotes-Genon:2014uoa} or even just by statistical fluctuations. In order to further understand the origin of the observed anomalies, it is very necessary to study other processes mediated by the same quark-level $b\to s\ell^+\ell^-$ transition.

In this respect, the rare baryonic $\Lambda_b\to \Lambda\mu^+\mu^-$ decay is of particular interest for the following two reasons. Firstly, due to the spin-half nature of $\Lambda_b$ and $\Lambda$ baryons, there is the potential to improve the currently limited understanding of the helicity structure of the underlying effective weak Hamiltonian~\cite{Mannel:1997xy,Chen:2001zc,Huang:1998ek}. Secondly, exploiting the full angular distribution of the four-body $\Lambda_b\to \Lambda(\to p\pi^-)\mu^+\mu^-$ decay, one can obtain information on the underlying short-distance Wilson coefficients of effective four-fermion operators, which is complementary to that obtained from the corresponding mesonic decays~\cite{Boer:2014kda,Gutsche:2013pp,Kumar:2015tnz}. Experimentally, this decay was observed firstly by the CDF collaboration with 24 signal events and a statistical significance of 5.8 Gaussian standard deviations~\cite{Aaltonen:2011qs}. Later, the LHCb collaboration published the first measurements of the differential branching fractions as well as three angular observables of this decay~\cite{Aaij:2013mna}. As the $\Lambda_b$ baryons account for around $20\%$ of the $b$-hadrons produced at the LHC~\cite{Aaij:2011jp}, refined measurements of this decay will be available in the near future. On the theoretical side, this decay is challenged by the hadronic uncertainties due to the $\Lambda_b\to \Lambda$ transition form factors and the non-factorizable spectator dynamics~\cite{Boer:2014kda,Feldmann:2011xf,Wang:2011uv,Wang:2015ndk}. As the theory of QCD factorization at low $q^2$~\cite{Beneke:2001at,Beneke:2004dp} is not yet fully developed for the baryonic decay, we neglect all the non-factorizable spectator-scattering effects. For the factorizable nonlocal hadronic matrix elements of the operators $O_{1-6,8}$, we absorb them into the effective Wilson coefficients $C_7^\mathrm{eff}(q^2)$ and $C_9^\mathrm{eff}(q^2)$~\cite{Beneke:2001at,Beneke:2004dp,Grinstein:2004vb,
Beylich:2011aq,Grinstein:1988me,Altmannshofer:2008dz}. For other previous studies of this decay, the readers are referred to Refs.~\cite{Chen:2002rg,Aliev:2002ww,Aliev:2002hj,Aliev:2002tr,
Aliev:2004af,Aliev:2004yf,Aliev:2005np,Giri:2005yt,Giri:2005mt,Turan:2005pf,
Turan:2005cw,Aliev:2006xd,Bashiry:2007pd,Zolfagharpour:2007eh,Wang:2008sm,
Aslam:2008hp,Wang:2008ni,Aliev:2010uy,Azizi:2010qk,Aliev:2012ac,Gan:2012tt,
Azizi:2013eta,Wen:2013ora,Liu:2015kaa,Mott:2015zma,Azizi:2015hoa,Sahoo:2016nvx,
Wang:2016dne}.

Interestingly, it has been observed by Meinel and Dyk~\cite{Meinel:2016grj} that the $\Lambda_b\to \Lambda(\to p\pi^-)\mu^+\mu^-$ decay prefers a positive shift to the Wilson coefficient $C_9$, which is opposite in sign compared to that found in the latest global fits of only mesonic decays~\cite{Altmannshofer:2014rta,Descotes-Genon:2015uva,Hurth:2016fbr}. This suggests that a simple shift in $C_9$ alone could not explain all the current data and needs more thorough analyses. In our previous paper~\cite{Hu:2016gpe}, we have studied the $B^0\to K^{\ast 0}\mu^+\mu^-$ decay in the aligned two-Higgs-doublet model (A2HDM)~\cite{Pich:2009sp}, and found that the angular observable $P_5'$ could be increased significantly to be consistent with the experimental data in the case when the charged-scalar contributions to $C_7^\mathrm{H^\pm}$ and $C_{9,10}^{\prime\mathrm{H^\pm}}$ are sizable, but $C_{9,10}^\mathrm{H^\pm}\simeq0$. In order to further understand the anomalies observed in the $b\to s\ell^+\ell^-$ mesonic decays, in this paper, we shall study the $\Lambda_b\to \Lambda(\to p\pi^-)\mu^+\mu^-$ decay in the A2HDM. As the $\Lambda_b$ polarization in the LHCb setup has been measured to be small and compatible with zero~\cite{Aaij:2013oxa}, and the polarization effect will be  averaged out for the symmetric ATLAS and CMS detectors, we consider only the case of unpolarized $\Lambda_b$ decay. In order to reduce as much as possible the uncertainties arising from input parameters and transition form factors, we shall calculate all of the angular observables in some appropriate combinations~\cite{Boer:2014kda,Gutsche:2013pp,Kumar:2015tnz}. For the $\Lambda_b\to \Lambda$ transition form factors, we use the latest high-precision lattice QCD calculation~\cite{Detmold:2016pkz}, which is extrapolated to the whole $q^2$ region using the Bourrely-Caprini-Lellouch parametrization~\cite{Bourrely:2008za}. These results are also consistent with those of the recent QCD light-cone sum rule calculation~\cite{Wang:2015ndk}, but with much smaller uncertainties in most of the kinematic range.

Our paper is organized as follows. In Sec.~\ref{sec:A2HDM}, we give a brief overview of the A2HDM. In Sec.~\ref{sec:calculate}, we present the theoretical framework for $\Lambda_b\to \Lambda(\to p\pi^-)\mu^+\mu^-$ decay, including the effective weak Hamiltonian, the $\Lambda_b\to\Lambda$ transition form factors, and the observables of this decay. In Sec.~\ref{sec:results}, we give our numerical results and discussions. Our conclusions are made in Sec.~\ref{sec:conclusion}. Some relevant formulae for the Wilson coefficients are collected in the appendix.

\section{The aligned two-Higgs-doublet model}
\label{sec:A2HDM}

We consider the minimal version of 2HDM, which is invariant under the SM gauge group and includes, besides the SM matter and gauge fields, two complex scalar $\mathrm{SU\left(2\right)_{L}}$ doublets, with hypercharge $Y=1/2$~\cite{Pich:2009sp,Branco:2011iw}. In the Higgs basis, the two doublets can be parametrized as
\begin{equation} \label{eq:Higgsbasis}
 \Phi_1=\left[ \begin{array}{c} G^+ \\ \frac{1}{\sqrt{2}}\, (v+S_1+iG^0) \end{array} \right] \; ,
 \qquad\qquad
 \Phi_2 = \left[ \begin{array}{c} H^+ \\ \frac{1}{\sqrt{2}}\, (S_2+iS_3)   \end{array}\right] \; ,
\end{equation}
where $v=(\sqrt{2} G_F)^{-1/2} \simeq 246~\mathrm{GeV}$ is the nonzero vacuum expectation value, and $G^\pm$, $G^0$ are the massless Goldstone fields. The remaining five physical degrees of freedom are given by the two charged fields $H^\pm(x)$ and the three neutral ones $\varphi^0_i(x) =\{h(x), H(x), A(x)\}=\mathcal{R}_{ij}S_j$, with the orthogonal transformation $\mathcal{R}$ fixed by the scalar potential~\cite{Pich:2009sp,Branco:2011iw,Li:2014fea}.

The most general Yukawa Lagrangian of the 2HDM is given by~\cite{Pich:2009sp}
\begin{equation}
 \mathcal{L}_Y = -\frac{\sqrt{2}}{v}\,\Big[\bar{Q}'_L (M'_d \Phi_1 + Y'_d \Phi_2) d'_R + \bar{Q}'_L (M'_u \tilde{\Phi}_1 + Y'_u \tilde{\Phi}_2) u'_R + \bar{L}'_L (M'_\ell \Phi_1 + Y'_\ell \Phi_2) \ell'_R \Big] + \mathrm{h.c.} \,,
\end{equation}
where $\tilde{\Phi}_a(x)\equiv i\tau_2\Phi_a^{\ast}(x)$ are the charge-conjugated scalar doublets with hypercharge $Y=-\frac{1}{2}$; $\bar{Q}_L'$ and $\bar{L}_L'$ are the left-handed quark and lepton doublets, and $u'_R$, $d'_R$ and $\ell'_R$ the corresponding right-handed singlets, in the weak-interaction basis. All fermionic fields are written as 3-dimensional vectors and the Yukawa couplings $M_f'$ and $Y_f'$~($f=u,d,\ell$) are therefore $3\times 3$ matrices in flavour space. Generally, the couplings $M_f'$ and $Y_f'$ can not be diagonalized simultaneously and the non-diagonal elements will give rise to unwanted tree-level FCNC interactions. In the fermion mass-eigenstate basis, with diagonal mass matrices $M_f$, the tree-level FCNCs can be eliminated by requiring the alignment in flavour space of the Yukawa matrices~\cite{Pich:2009sp}:
\begin{equation}
 Y_{d,\ell}=\varsigma_{d,\ell}\, M_{d,\ell}\, ,
 \qquad
 Y_u=\varsigma^\ast_u\, M_u\, ,
\end{equation}
where $\varsigma_f$~($f=u,d,\ell$) are arbitrary complex parameters and could introduce new sources of CP violation beyond the SM.

The interactions of the charged scalars with the fermion mass-eigenstate fields read~\cite{Pich:2009sp}
\begin{equation}
 \mathcal{L}_{H^\pm}=- \frac{\sqrt{2}}{v}\, H^+\, \Big\{\bar{u} \left[\varsigma_d\,V\, M_d P_R - \varsigma_u\,M_u^{\dagger} V\, P_L\right] d + \varsigma_\ell\,\bar{\nu} M_\ell P_R \ell \Big\}+ \mathrm{h.c.} \,,
\end{equation}
where $P_{R(L)}\equiv \frac{1\pm\gamma_5}{2}$ is the right~(left)-handed chirality projector, and $V$ denotes the Cabibbo-Kobayashi-Maskawa (CKM) matrix~\cite{Cabibbo:1963yz,Kobayashi:1973fv}. As detailed in Refs.~\cite{Li:2014fea,Hu:2016gpe}, the charged scalars could provide large contributions to $b\to s\ell^+\ell^-$ transitions, in some given parameter spaces.

\section{The $\boldsymbol{\Lambda_b\to \Lambda(\to p\pi^-)\mu^+\mu^-}$ decay}
\label{sec:calculate}

\subsection{Effective weak Hamiltonian}
\label{subsec:Heff}

The effective weak Hamiltonian for $b\to s\ell^+\ell^-$ transition is given by~\cite{Altmannshofer:2008dz}
\begin{eqnarray} \label{eq:Heff}
\mathcal{H}_{\mathrm{eff}}=-\frac{4G_{F}}{\sqrt{2}}V_{tb}V_{ts}^{\ast}
\left[C_1O_1^c+C_2O_2^c+\sum_{i=3}^{10}C_iO_i+\sum_{i=7,9,10}C_{i}'O_{i}'\right]\,,
\end{eqnarray}
where $G_F$ is the Fermi coupling constant, and we have neglected the doubly Cabibbo-suppressed contributions to the decay amplitude. The operators $O_{i\leq6}$ are identical to $P_i$ given in Ref.~\cite{Bobeth:1999mk}, and the remaining ones read
\begin{align}
O_7^{(\prime)} &= \frac{e}{16\pi^2}\bar m_b\left(\bar s\sigma^{\mu\nu}P_{R(L)} b\right)F_{\mu\nu}\,, &
O_8 &= \frac{g_s}{16\pi^2}\bar m_b\left(\bar s\sigma^{\mu\nu}T^a P_R b\right)G^a_{\mu\nu}\,,
\\[0.2cm]	
O_9^{(\prime)} &= \frac{e^2}{16\pi^2}\left(\bar s\gamma^\mu P_{L(R)} b\right)\left(\bar \ell\gamma_\mu \ell\right)\,, &
O_{10}^{(\prime)} &= \frac{e^2}{16\pi^2}\left(\bar s\gamma^\mu P_{L(R)} b\right)\left(\bar\ell\gamma_\mu\gamma_5 \ell\right)\,,
\end{align}
where $e\,(g_s)$ is the electromagnetic~(strong) coupling constant, $T^a$ the generator of $\mathrm{SU(3)_C}$ in the fundamental representation, and $\bar m_b$ denotes the $b$-quark running mass in the $\mathrm{\overline{MS}}$ scheme.

Within the SM, $O_{7,9,10}$ play the leading role in $b\to s\ell^+\ell^-$ transition, while the factorizable contributions from $O_{1-6,8}$ can be absorbed into the effective Wilson coefficients $C_7^\mathrm{eff}(q^2)$ and $C_9^\mathrm{eff}(q^2)$~\cite{Du:2015tda}:
\begin{align}
C_7^{\rm eff}(q^2) &= C_7 - \frac{1}{3} \left( C_3 + \frac{4}{3}\,C_4 + 20\,C_5  + \frac{80}{3}\, C_6 \right) - \frac{\alpha_s}{4 \pi} \left[ \left(C_1 - 6\,C_2\right) F_{1,c}^{(7)}(q^2) + C_8\, F_8^{(7)}(q^2) \right], \label{eq:C7eff} \\
C_9^{\rm eff}(q^2) &= C_9 + \frac{4}{3}\left( C_3 + \frac{16}{3}\, C_5 + \frac{16}{9}\, C_6\right)- h(0,q^2) \left( \frac{1}{2}\, C_3 + \frac{2}{3}\, C_4 + 8\, C_5 + \frac{32}{3}\, C_6 \right) \nonumber \\[0.1cm]
& - h(m_b,q^2) \left( \frac{7}{2}\, C_3 + \frac{2}{3}\, C_4 + 38\, C_5 + \frac{32}{3}\, C_6\right)+ h(m_c,q^2) \left( \frac{4}{3}\, C_1 + C_2 + 6\, C_3 + 60\, C_5 \right) \nonumber \\[0.1cm]
& - \frac{\alpha_s}{4 \pi} \bigg[ C_1\, F_{1,c}^{(9)}(q^2) + C_2\, F_{2,c}^{(9)}(q^2) + C_8\, F_8^{(9)}(q^2) \bigg]\,, \label{eq:C9eff}
\end{align}
where the basic fermion loop function is given by~\cite{Beneke:2001at}
\begin{align}
h(m_q,q^2) =\;  & \frac{4}{9} \left(\ln \frac{\mu^2}{m_q^2} + \frac{2}{3} + z\right) \nonumber \\
 & -\frac{4}{9} (2+z) \sqrt{|z-1|}
	\begin{cases}
		\arctan \frac{1}{\sqrt{z-1}}\,, & z= \frac{4m_q^2}{q^2} > 1 \cr
		\ln \frac{1+ \sqrt{1-z}}{\sqrt{z}} -\frac{i \pi}{2}\,, & z = \frac{4m_q^2}{q^2}\leq 1 \cr
	\end{cases}\,,
\end{align}
and the functions $F_8^{(7,9)}(q^2)$ are defined by Eqs.~(B.1) and (B.2) of Ref.~\cite{Beneke:2001at}, while $F_{1,c}^{(7,9)}(q^2)$ and $F_{2,c}^{(7,9)}(q^2)$ are provided in Ref.~\cite{Asatryan:2001zw} for low $q^2$ and in Ref.~\cite{Greub:2008cy} for high $q^2$.\footnote{Here we incorporate only the leading contributions from an operator product expansion (OPE) of the nonlocal product of $O_{1-6,8}$ with the quark electromagnetic current, because the first and second-order corrections in $\Lambda/m_b$ from the OPE are already well suppressed in the high-$q^2$ region~\cite{Grinstein:2004vb,Beylich:2011aq}. Although non-factorizable spectator-scattering effects (\emph{i.e.}, corrections that are not described using hadronic form factors) are expected to play a sizable role in the low-$q^2$ region~\cite{Beneke:2001at,Beneke:2004dp}, we shall neglect their contributions because there is presently no systematic framework in which they can be calculated for the baryonic decay~\cite{Wang:2015ndk}. As a consequence, our predictions in the low-$q^2$ region are affected by a hitherto unquantified systematic uncertainty.} The quark masses appearing in these functions are defined in the pole scheme. The contribution from $O_7'$ is suppressed by $\bar{m}_s/\bar{m}_b$ and those from $O_{9,10}'$ are zero within the SM.

%%%%%%%%%%%%%%%%%%%%%%%%%%%%%%%%%%%%%%%%%%%%%%%%%%%%%%%
\begin{figure}[t]
  \centering
  \includegraphics[width=0.85\textwidth]{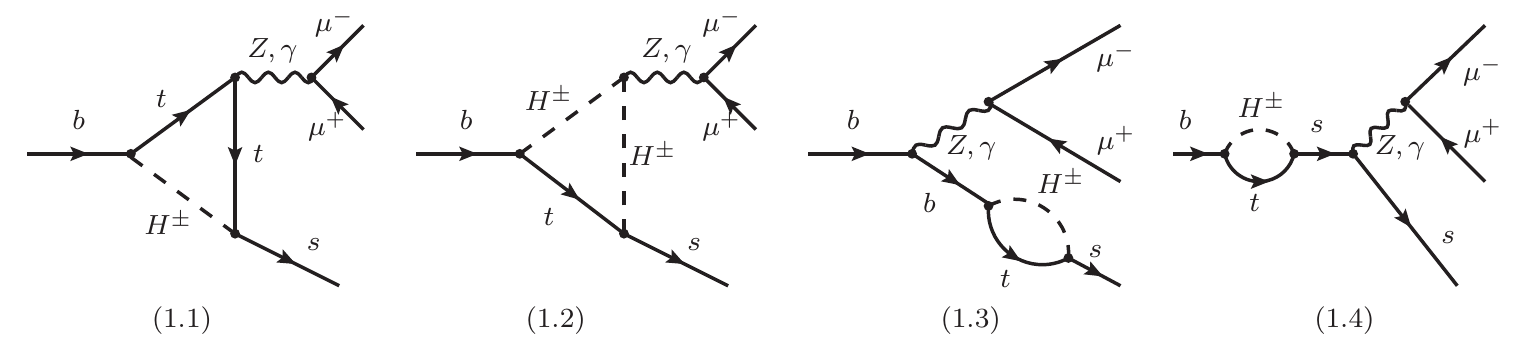}
  \caption{\small $Z$- and photon-penguin diagrams involving the charged-scalar exchanges in the A2HDM.}
  \label{fig:ZgA2HDM}
\end{figure}
%%%%%%%%%%%%%%%%%%%%%%%%%%%%%%%%%%%%%%%%%%%%%%%%%%%%%%%

In the A2HDM, the charged-scalar exchanges lead to additional contributions to $C_{7,9,10}$ and make the contributions of chirality-flipped operators $O'_{7,9,10}$ to be significant, through the $Z^0$- and photon-penguin diagrams shown in Figure~\ref{fig:ZgA2HDM}. Since we have neglected the light lepton mass, there is no contribution from the SM $W$-box diagrams with the $W^{\pm}$ bosons replaced by the charged scalars $H^{\pm}$. The new contributions to the Wilson coefficients read~\cite{Hu:2016gpe}
\begin{align}\label{eq:WCinA2HDM}
&C_{7}^\mathrm{H^\pm}=\left|\varsigma_{u}\right|^{2}C_{7,\,\mathrm{uu}} +\varsigma_{d}\varsigma_{u}^{\ast}C_{7,\,\mathrm{ud}}\,,\\[0.2cm]
&C_{9}^\mathrm{H^\pm}=\left|\varsigma_{u}\right|^{2}C_{9,\,\mathrm{uu}}\,,\\[0.2cm]
&C_{10}^\mathrm{H^\pm}=\left|\varsigma_{u}\right|^{2}C_{10,\,\mathrm{uu}}\,,\\[0.2cm]
&C_{7}^{\prime\mathrm{H^\pm}}=\frac{\bar m_{s}}{\bar m_{b}}\Big[\left|\varsigma_{u}\right|^{2} C_{7,\,\mathrm{uu}} +\varsigma_{u}\varsigma_{d}^{\ast}C_{7,\,\mathrm{ud}}\Big]\,,\\[0.2cm]
&C_{9}^{\prime\mathrm{H^\pm}}=\left(-1+4\sin^{2}\theta_{W}\right)
C_{10}^{\prime\mathrm{H^\pm}}+\frac{\bar m_{b}\bar m_{s}}{m_{W}^{2}}
\Big[\left|\varsigma_{u}\right|^{2}C_{9,\,\mathrm{uu}}'
+2\Re\left(\varsigma_{u}\varsigma_{d}^{\ast}\right)C_{9,\,\mathrm{ud}}'
+\left|\varsigma_{d}\right|^{2}C_{9,\,\mathrm{dd}}'\Big]\,,\\[0.2cm]
&C_{10}^{\prime\mathrm{H^\pm}}=\frac{\bar m_{b}\bar m_{s}}{m_{W}^{2}}
\Big[\left|\varsigma_{u}\right|^{2}C_{10,\,\mathrm{uu}}'
+2\Re\left(\varsigma_{u}\varsigma_{d}^{\ast}\right)C_{10,\,\mathrm{ud}}'
+\left|\varsigma_{d}\right|^{2}C_{10,\,\mathrm{dd}}'\Big]\,,
\end{align}
with the functions $C_{i,{\rm XY}}^{(\prime)}$~($i=7,9,10$; ${\rm X,Y=u,d}$) given by Eqs.~(\ref{eq:WCA2HDM1})--(\ref{eq:WCA2HDM10}). Assuming $\varsigma_{u,d}$ to be real, one has $C_{7}^{\prime\mathrm{H^\pm}}=\frac{\bar m_{s}}{\bar m_{b}}C_{7}^\mathrm{H^\pm}$, and we shall therefore neglect $C_{7}^{\prime\mathrm{H^\pm}}$in the following discussion.

\subsection{Transition form factors}
\label{subsec:TFFs}

In order to obtain compact forms of the helicity amplitudes~\cite{Boer:2014kda}, we adopt the helicity-based definitions of the $\Lambda_b\to\Lambda$ transition form factors, which are given by~\cite{Boer:2014kda,Feldmann:2011xf}
\begin{align}
\langle\Lambda(p^\prime,s^\prime)|\bar{s}\,\gamma^\mu\,b|\Lambda_b(p,s)\rangle =\; & \bar{u}_\Lambda(p^\prime,s^\prime) \bigg[\,f_t^V(q^2)\, (m_{\Lambda_b}-m_\Lambda)\,\frac{q^\mu}{q^2} \nonumber \\[0.2cm]
& + f_0^V(q^2)\,\frac{m_{\Lambda_b}+m_\Lambda}{s_+}\, \Big(p^\mu+p^{\prime\mu}-(m_{\Lambda_b}^2-m_\Lambda^2)\,
\frac{q^\mu}{q^2}\Big) \nonumber \\[0.2cm]
& + f_\perp^V(q^2)\,\Big(\gamma^\mu-\frac{2m_\Lambda}{s_+}\, p^\mu-\frac{2m_{\Lambda_b}}{s_+}\,p^{\prime\mu}\Big)\,\bigg]\, u_{\Lambda_b}(p,s)\,,\\[0.3cm]
\langle\Lambda(p^\prime,s^\prime)|\bar{s}\,\gamma^\mu\gamma_5\,b| \Lambda_b(p,s)\rangle =\; & -\bar{u}_\Lambda(p^\prime,s^\prime)\,\gamma_5\, \bigg[\,f_t^A(q^2)\, (m_{\Lambda_b}+m_\Lambda)\,\frac{q^\mu}{q^2} \nonumber \\[0.2cm]
& + f_0^A(q^2)\,\frac{m_{\Lambda_b}-m_\Lambda}{s_-}\, \Big(p^\mu+p^{\prime\mu}- (m_{\Lambda_b}^2-m_\Lambda^2)\,\frac{q^\mu}{q^2}\Big) \nonumber\\[0.2cm]
& + f_\perp^A(q^2)\,\Big(\gamma^\mu+\frac{2m_\Lambda}{s_-}\,p^\mu -\frac{2m_{\Lambda_b}}{s_-}\,p^{\prime\mu}\Big)\,\bigg]\,u_{\Lambda_b}(p,s)\,,
\end{align}
for the vector and axial-vector currents, respectively, and
\begin{align}
\langle\Lambda(p^\prime,s^\prime)|\bar{s}\,i\sigma^{\mu\nu}q_\nu\,b| \Lambda_b(p,s)\rangle = \; & -\bar{u}_\Lambda(p^\prime,s^\prime)\,\bigg[\,f_0^T(q^2)\,\frac{q^2}{s_+}\, \Big(p^\mu+p^{\prime\mu}-(m_{\Lambda_b}^2-m_{\Lambda}^2)\,\frac{q^\mu}{q^2} \Big) \nonumber\\[0.2cm]
& + f_\perp^T(q^2)\,(m_{\Lambda_b}+m_\Lambda)\, \Big(\gamma^\mu-\frac{2m_\Lambda}{s_+}\,p^\mu-\frac{2m_{\Lambda_b}}{s_+}\, p^{\prime\mu}\Big)\,\bigg]\,u_{\Lambda_b}(p,s)\,,\\[0.3cm]
\langle\Lambda(p^\prime,s^\prime)|\bar{s}\,i\sigma^{\mu\nu}\gamma_5\,q_\nu\,b| \Lambda_b(p,s)\rangle =\; & - \bar{u}_{\Lambda}(p^\prime,s^\prime)\, \gamma_5\,\bigg[\,f_0^{T_5}(q^2)\,\frac{q^2}{s_-}\, \Big(p^\mu+p^{\prime\mu}-(m_{\Lambda_b}^2-m_{\Lambda}^2)\,\frac{q^\mu}{q^2} \Big)\nonumber\\[0.2cm]
& + f_\perp^{T_5}(q^2)\,(m_{\Lambda_b}-m_\Lambda)\, \Big(\gamma^\mu+\frac{2m_\Lambda}{s_-}\,p^\mu-\frac{2m_{\Lambda_b}}{s_-}\, p^{\prime\mu}\Big)\,\bigg]\,u_{\Lambda_b}(p,s)\,,
\end{align}
for the tensor and pseudo-tensor currents, respectively. Here $q=p-p'$ and $s_\pm=(m_{\Lambda_b}\pm m_\Lambda)^2-q^2$. The helicity form factors satisfy the endpoint relations $f_t^{V(A)}(0)=f_0^{V(A)}(0)$ and $f_\perp^{A(T_5)}(q_{\rm max}^2)=f_0^{A(T_5)}(q_{\rm max}^2)$, with $q^2_{\rm max}=(m_{\Lambda_b}-m_\Lambda)^2$. All these ten form factors have been recently calculated using $(2+1)$-flavour lattice QCD~\cite{Detmold:2016pkz}.

\subsection{Observables in $\boldsymbol{\Lambda_b\to\Lambda(\to p\pi^-)\mu^+\mu^-}$ decay}

The angular distribution of the four-body $\Lambda_b\to\Lambda(\to p\pi^-)\mu^+\mu^-$ decay, with an unpolarized $\Lambda_b$, is described by the dimuon invariant mass squared $q^2$, the helicity angles $\theta_\Lambda$ and $\theta_\ell$, and the azimuthal angle $\phi$; the explicit definition of these four kinematic variables could be found, for example, in Refs.~\cite{Boer:2014kda,Gutsche:2013pp}. The four-fold differential width can then be written as~\cite{Boer:2014kda}
\begin{align}\label{eq:4-fold}
\frac{\mathrm{d}^4 \Gamma}{\mathrm{d} q^2\, \mathrm{d}\cos\theta_\ell\, \mathrm{d}\cos\theta_\Lambda\, \mathrm{d} \phi} \;=\;
&\frac{3}{8\pi}\,\bigg[\,
\Big(K_{1ss}\sin^2\theta_\ell+\,K_{1cc}\cos^2\theta_\ell+K_{1c}\cos\theta_\ell\Big) \nonumber \\[0.2cm]
&\quad+\Big(K_{2ss}\sin^2\theta_\ell+\,K_{2cc}\cos^2\theta_\ell+K_{2c}
\cos\theta_\ell\Big)\cos\theta_\Lambda \nonumber \\[0.2cm]
&\quad+\Big(K_{3sc}\sin\theta_\ell\cos\theta_\ell+K_{3s}\sin\theta_\ell\Big)\,
\sin\theta_\Lambda\,\cos\phi \nonumber \\[0.2cm]
&\quad+\Big(K_{4sc}\sin\theta_\ell\cos\theta_\ell+K_{4s}\sin\theta_\ell\Big)\,
\sin\theta_\Lambda\,\sin\phi\,\bigg],
\end{align}
where the angular coefficients $K_{n\lambda}$, with $n=1,\dots,4$ and $\lambda=s,c,ss,cc,sc$, are functions of $q^2$, and can be expressed in terms of eight transversity amplitudes for $\Lambda_b\to\Lambda$ transition and the parity-violating decay parameter $\alpha_{\Lambda}$ in the secondary decay $\Lambda\to p\pi^-$; their explicit expressions could be found in Ref.~\cite{Boer:2014kda}.

Starting with Eq.~\eqref{eq:4-fold} and in terms of the angular coefficients $K_{n\lambda}$, we can then construct the following observables~\cite{Boer:2014kda,Gutsche:2013pp,Kumar:2015tnz}:
\begin{itemize}
\item The differential decay rate and differential branching fraction
  \begin{align}\label{eq:branching}
  \frac{{\rm d}\Gamma}{{\rm d}q^2}=2K_{1ss}+K_{1cc}\,, \qquad
  \frac{{\rm d}{\cal B}}{{\rm d}q^2}=\tau_{\Lambda_b}
  \frac{{\rm d}\Gamma}{{\rm d}q^2}\,,
  \end{align}
  where $\tau_{\Lambda_b}$ is the $\Lambda_b$ lifetime.

\item The longitudinal polarization fraction of the dimuon system
  \begin{align}\label{eq:longitudinally}
  F_L=2\hat K_{1ss}-\hat K_{1cc}\,,
  \end{align}
  where we introduce the normalized angular observables $\hat K_{n\lambda}=\frac{K_{n\lambda}}{{\rm d}\Gamma/{\rm d}q^2}$.

\item The lepton-, hadron- and combined lepton-hadron-side forward-backward asymmetries
  \begin{align}
  A_{\rm FB}^\ell=\frac{3}{2}\hat K_{1c}\,,\qquad
  A_{\rm FB}^\Lambda=\hat K_{2ss}+\frac{1}{2}\hat K_{2cc}\,,\qquad
  A_{\rm FB}^{\ell\Lambda}=\frac{3}{4}\hat K_{2c}\,,
  \end{align}
  which have characteristic $q^2$ behaviours: within the SM, both $A_{\rm FB}^\ell$ and $A_{\rm FB}^{\ell\Lambda}$ have the same zero-crossing points, $q^2_0(A_{\rm FB}^\ell)\simeq q^2_0(A_{\rm FB}^{\ell\Lambda})$, to the first approximation, while $A_{\rm FB}^\Lambda$ does not cross zero~\cite{Boer:2014kda}. Note that, as observed in the mesonic case~\cite{Beneke:2001at,Ali:1991is,Burdman:1998mk,Ali:1999mm}, the zero-crossing points are nearly free of hadronic uncertainties~\cite{Boer:2014kda,Gutsche:2013pp,Kumar:2015tnz}.

\item The other five asymmetry observables
  \begin{align}
  & Y_2=\frac{3}{8}(\hat K_{2cc}-\hat K_{2ss})\,, \qquad
    Y_{\rm3\,s}=\frac{3\pi}{8}\hat K_{3s}\,, \qquad
    Y_{\rm3\,sc}=\frac{1}{2}\hat K_{3sc}\,, \nonumber \\[0.2cm]
  & Y_{\rm4\,s}=\frac{3\pi}{8}\hat K_{4s}\,, \qquad
    Y_{\rm4\,sc}=\frac{1}{2}\hat K_{4sc}\,,\label{eq:Y4sc}
  \end{align}
  which, along with the previous observables, determine all the ten angular coefficients $K_{n\lambda}$. Here $Y_2$ also has a zero-crossing point, which lies in the low $q^2$ region.
\end{itemize}
In order to compare with the experimental data~\cite{Aaij:2015xza}, we also consider the binned differential branching fraction defined by
\begin{eqnarray}
\langle {\rm d}{\cal B}/{\rm d}q^2 \rangle_{[q^2_{\rm min},q^2_{\rm max}]} =\frac{\displaystyle\int_{q^2_{\rm min}}^{q^2_{\rm max}}\left({\rm d}{\cal B}/{\rm d}q^2\right){\rm d}q^2}{q^2_{\rm max}-q^2_{\rm min}}\,,
\end{eqnarray}
and the binned normalized angular coefficients defined by
\begin{eqnarray}
\langle \hat K_{n\lambda} \rangle_{[q^2_{\rm min},q^2_{\rm max}]} =\frac{\displaystyle\int_{q^2_{\rm min}}^{q^2_{\rm max}}K_{n\lambda}\;{\rm d}q^2}{\displaystyle\int_{q^2_{\rm min}}^{q^2_{\rm max}}\left({\rm d}\Gamma/{\rm d}q^2\right){\rm d}q^2}\,,
\end{eqnarray}
where the numerator and denominator should be binned separately. As the theoretical calculations are thought to break down close to the narrow charmonium resonances, we make no predictions for these observables in this region.

Finally, it should be noted that, unlike the strong decay $K^\ast\to K\pi$ in the mesonic counterpart $B\to K^\ast\ell^+\ell^-$, the subsequent weak decay $\Lambda\to p\pi^-$ is parity violating, with the asymmetry parameter $\alpha_{\Lambda}$ being known from experiment~\cite{Olive:2016xmw}. This fact makes the signal with an intermediate $\Lambda$ baryon to be distinguished from the direct $\Lambda_b\to p\pi^-\mu^+\mu^-$ decay, and facilitates the full angular analysis of $\Lambda_b\to\Lambda(\to p\pi^-)\mu^+\mu^-$ decay~\cite{Boer:2014kda,Gutsche:2013pp}.

\section{Numerical results and discussions}
\label{sec:results}

\subsection{Input parameters}

Firstly we collect in Table~\ref{tab:inputs} the theoretical input parameters entering our numerical analysis throughout this paper. These include the SM parameters such as the electromagnetic and strong coupling constants, gauge boson, quark and hadron masses\footnote{The pion mass is needed to describe the secondary decay $\Lambda\to p\pi^-$, and the kaon and $B$-meson masses are used to evaluate the $BK$ threshold in the $z$ parametrization of the $\Lambda_b\to \Lambda$ transition form factors~\cite{Detmold:2016pkz}.}, as well as the CKM matrix elements. The weak mixing angle $\theta_W$ is given by $\sin^2\theta_W=1-M_W^2/M_Z^2$.

%%%%%%%%%%%%%%%%%%%%%%%%%%%%%%%%%%%%%%%%%%%%%%%%%%%%%%%%%%%%%%%%%%%
\begin{table}[t]
\begin{center}
\caption{\label{tab:inputs} Summary of the theoretical input parameters used throughout this paper.}
\vspace*{0.3cm}
\renewcommand{\arraystretch}{1.0}
{\tabcolsep=0.407cm\begin{tabular}{|ccccc|c|}
\hline\hline
\multicolumn{6}{|l|}{\hspace{0.5cm}\textbf{QCD and electroweak parameters}}
\\
\hline
  $G_F [10^{-5}\gev^{-2}]$
& $\alpha_s(m_Z)$
& $\alpha_e(m_W)$
& $m_Z [\gev]$
& $m_W [\gev]$
& \cite{Olive:2016xmw}
\\
  $1.1663787$
& $0.1182 \pm 0.0012$
& $1/128$
& $91.1876$
& $80.385$
&
\\
\hline
\end{tabular}}
{\tabcolsep=0.256cm \begin{tabular}{|cccccc|c|}
\hline
\multicolumn{7}{|l|}{\hspace{0.65cm}\textbf{Quark masses [GeV]}}
\\
\hline
  $m_t^{\rm pole}$
& $m_b^{\rm pole}$
& $m_c^{\rm pole}$
& $\overline{m}_b(\overline{m}_b)$
& $\overline{m}_c(\overline{m}_c)$
& $\overline{m}_s(2{\rm GeV})$
& \cite{Olive:2016xmw}
\\
  $174.2 \pm 1.4$
& $4.78 \pm 0.06$
& $1.67 \pm 0.07$
& $4.18  \pm 0.03$
& $1.27 \pm 0.03$
& $0.096_{-0.004}^{+0.008}$
&
\\
\hline
\end{tabular}}
{\tabcolsep=0.834cm \begin{tabular}{|ccccc|c|}
\hline
\multicolumn{6}{|l|}{\textbf{Meson and baryon masses [GeV]}}
\\
\hline
  $m_\pi$
& $m_K$
& $m_B$
& $m_\Lambda$
& $m_{\Lambda_b}$
& \cite{Detmold:2016pkz,Olive:2016xmw}
\\
  $0.135$
& $0.494$
& $5.279$
& $1.116$
& $5.619$
&
\\
\hline
\end{tabular}}
{\tabcolsep=0.316cm\begin{tabular}{|ccc|c|}
\hline
\multicolumn{4}{|l|}{\hspace{0.65cm}\textbf{Other parameters}}
\\
\hline
  $\Lambda_b$ lifetime($\tau_{\Lambda_b}$)
& parity-violating parameter ($\alpha_{\Lambda}$)
& $|V_{ts}^\ast V_{tb}|$
& \cite{Olive:2016xmw,UTfit:2016}
\\
  $(1.466\pm0.010)$ ps
& $0.642\pm0.013$
& $0.04152\pm0.00056$
&
\\
\hline
\end{tabular}}
\renewcommand{\arraystretch}{1.0}
\end{center}
\end{table}
%%%%%%%%%%%%%%%%%%%%%%%%%%%%%%%%%%%%%%%%%%%%%%%%%%%%%%%%%%%%%%%%%%%

For the $\Lambda_b\to \Lambda$ transition form factors, we use the latest
high-precision lattice QCD calculation with $2+1$ dynamical flavours~\cite{Detmold:2016pkz}. The $q^2$ dependence of these form factors are parametrized in a simplified $z$ expansion~\cite{Bourrely:2008za}, modified to account for pion-mass and lattice-spacing dependences. All relevant formulae and input parameters can be found in Eqs.~(38) and (49) and Tables~III--V and IX--XII of Ref.~\cite{Detmold:2016pkz}. To compute the central value, statistical uncertainty, and total systematic uncertainty of any observable depending on the form factors, such as the differential branching fraction and angular observables given in Eqs.~\eqref{eq:branching}--\eqref{eq:Y4sc}, as well as the corresponding binned observables and the zero-crossing points, we follow the same procedure as specified in Eqs.~(50)--(55) of Ref.~\cite{Detmold:2016pkz}.

\subsection{Results within the SM}

For the short-distance Wilson coefficients at the low scale $\mu_b=4.2~{\rm GeV}$, we use the numerical values collected in Table~\ref{tab:WCiSM}, which are obtained at the next-to-next-to-leading logarithmic (NNLL) accuracy within the SM~\cite{Blake:2016olu,Bobeth:1999mk,Gambino:2003zm,Misiak:2004ew,Gorbahn:2004my,favio}.

%%%%%%%%%%%%%%%%%%%%%%%%%%%%%%%%%%%%%%%%%%%%%%%%%%%%%%%%%%%%%%%%%%%
\begin{table}[t]
\begin{center}
\caption{\label{tab:WCiSM}The Wilson coefficients at the scale $\mu_b=4.2~{\rm GeV}$, to NNLL accuracy in the SM.}
\vspace*{0.3cm} \tabcolsep 0.08in
\renewcommand{\arraystretch}{1.0}
\begin{tabular}{|cccccccccc|}
\hline\hline
  $C_1$
& $C_2$
& $C_3$
& $C_4$
& $C_5$
& $C_6$
& $C_7$
& $C_8$
& $C_9$
& $C_{10}$
\\
\hline
  $-0.294$
& $1.017$
& $-0.0059$
& $-0.087$
& $0.0004$
& $0.0011$
& $-0.324$
& $-0.176$
& $4.114$
& $-4.193$
\\
\hline
\end{tabular}
\end{center}
\end{table}
%%%%%%%%%%%%%%%%%%%%%%%%%%%%%%%%%%%%%%%%%%%%%%%%%%%%%%%%%%%%%%%%%%%

%%%%%%%%%%%%%%%%%%%%%%%%%%%%%%%%%%%%%%%%%%%%%%%%%%%%%%%
\begin{figure}[htbp]
  \centering
  \includegraphics[width=0.76\textwidth]{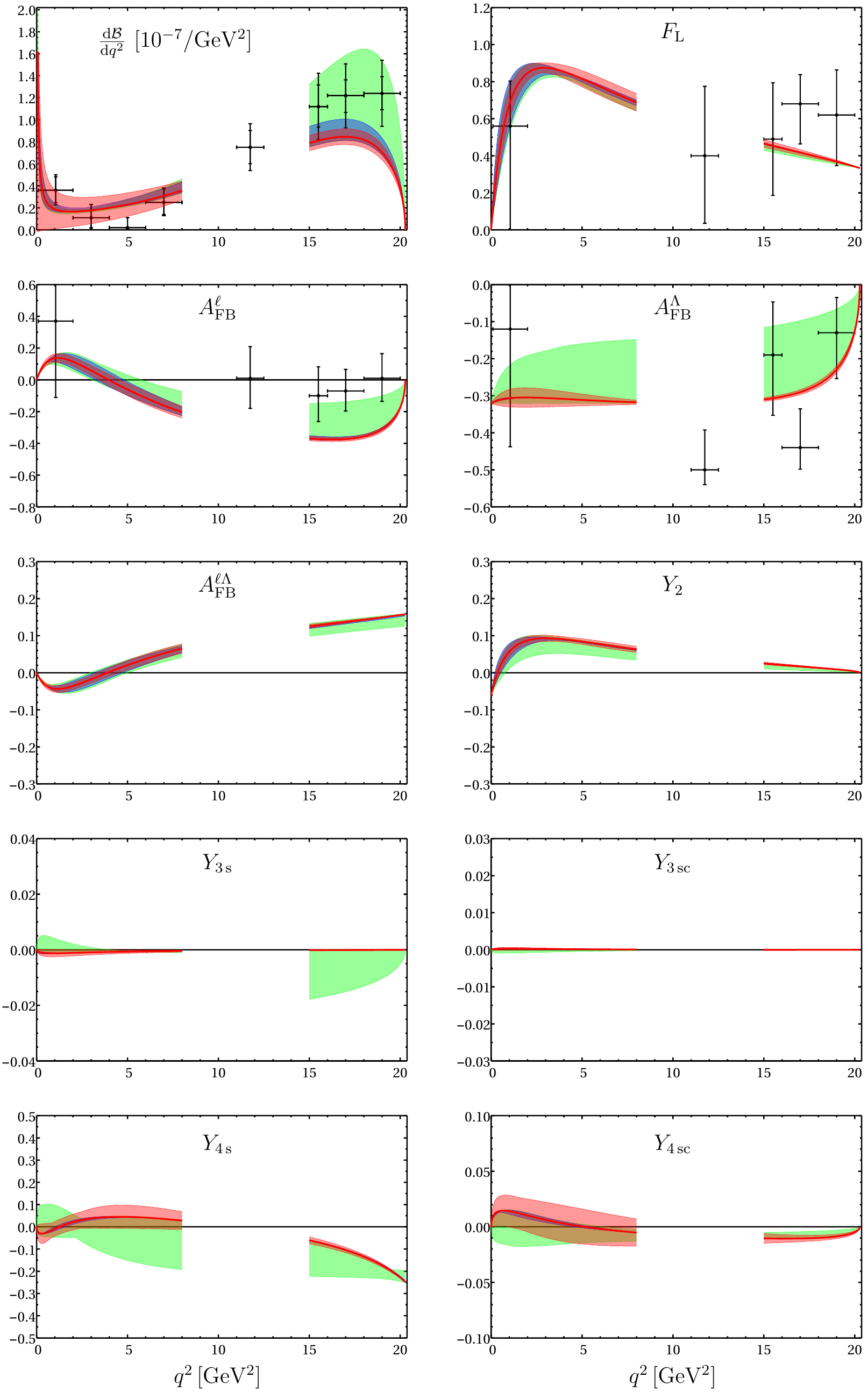}
  \caption{\small The $\Lambda_b\to\Lambda(\to p\pi^-)\mu^+\mu^-$ observables as a function of the dimuon invariant mass squared $q^2$, predicted both within the SM (central values: red solid curves, theoretical uncertainties: red bands) and in the A2HDM (case~A: blue bands and case~B: green bands). The corresponding experimental data from LHCb~\cite{Aaij:2015xza}, where available, are represented by the error bars.}
  \label{fig:a2hdm}
\end{figure}
%%%%%%%%%%%%%%%%%%%%%%%%%%%%%%%%%%%%%%%%%%%%%%%%%%%%%%%

We show in Figure~\ref{fig:a2hdm} the SM predictions for the differential branching fraction and angular observables as a function of the dimuon invariant mass squared $q^2$, where the central values are plotted as red solid curves and the theoretical uncertainties, which are caused mainly by the $\Lambda_b\to\Lambda$ transform form factors, are labelled by the red bands. The latest experimental data from LHCb~\cite{Aaij:2015xza}, where available, are also included in the figure for comparison\footnote{For the differential branching fraction, the error bars are shown both including and excluding the uncertainty from the normalization mode $\Lambda_b\to J/\psi\Lambda$~\cite{Olive:2016xmw}.}. The SM predictions for the corresponding binned observables are presented in Table~\ref{tab:binSM}.

%%%%%%%%%%%%%%%%%%%%%%%%%%%%%%%%%%%%%%%%%%%%%%%%%%%%%%%
\begin{table}[t]
\begin{center}
\caption{\label{tab:binSM}SM predictions for the binned differential branching fraction (in units of $10^{-7}~{\rm GeV}^{-2}$) and angular observables. The first column specifies the bin ranges $[q^2_{\rm min},q^2_{\rm max}]$ in units of ${\rm GeV}^2$. }
\vspace*{0.3cm} \tabcolsep 0.12in
\renewcommand{\arraystretch}{1.0}
\begin{tabular}{rccccc}
\hline \hline
& $\langle \mathrm{d}\mathcal{B}/\mathrm{d} q^2\rangle$
& $\langle F_L \rangle $
& $\langle A_{\rm FB}^\ell \rangle $
& $\langle A_{\rm FB}^\Lambda \rangle $
& $\langle A_{\rm FB}^{\ell\Lambda} \rangle $ \\
\hline
$[0.1,2]$ & $0.24(23)$  & $0.531(92)$  & $0.104(17)$  & $-0.310(18)$  & $-0.0332(56)$ \\
$[2,4]$   & $0.18(12)$  & $0.867(28)$  & $0.056(31)$  & $-0.306(24)$  & $-0.017(10)$ \\
$[4,6]$   & $0.23(11)$  & $0.813(43)$  & $-0.063(39)$ & $-0.311(16)$  & $0.021(13)$ \\
$[6,8]$   & $0.310(94)$ & $0.728(48)$  & $-0.163(39)$ & $-0.3160(88)$ & $0.053(13)$ \\
$[1.1,6]$ & $0.20(12)$  & $0.829(33)$  & $0.011(31)$  & $-0.309(20)$  & $-0.0026(99)$ \\
$[15,16]$ & $0.810(70)$ & $0.454(20)$  & $-0.373(14)$ & $-0.3069(55)$ & $0.1283(48)$ \\
$[16,18]$ & $0.839(72)$ & $0.418(15)$  & $-0.371(13)$ & $-0.2890(69)$ & $0.1372(37)$ \\
$[18,20]$ & $0.671(65)$ & $0.3711(78)$ & $-0.308(15)$ & $-0.2266(91)$ & $0.1487(19)$ \\
$[15,20]$ & $0.766(69)$ & $0.409(13)$  & $-0.349(13)$ & $-0.2709(73)$ & $0.1394(32)$ \\
\hline \hline
& $\langle Y_2 \rangle $
& $\langle Y_{\rm3\,s} \rangle\times10^{-2} $
& $\langle Y_{\rm3\,sc} \rangle\times10^{-3} $
& $\langle Y_{\rm4\,s} \rangle $
& $\langle Y_{\rm4\,sc} \rangle $ \\
\hline
$[0.1,2]$ & $0.032(17) $ & $-0.11(12)$  & $0.25(33)$   & $-0.013(29)$  & $0.013(12)$ \\
$[2,4]$   & $0.0918(84)$ & $-0.093(92)$ & $0.22(24)$   & $0.036(43)$   & $0.006(16)$ \\
$[4,6]$   & $0.0838(76)$ & $-0.066(62)$ & $0.15(14)$   & $0.044(52)$   & $0.000(16)$ \\
$[6,8]$   & $0.0696(85)$ & $-0.048(44)$ & $0.100(98)$  & $0.035(45)$   & $-0.004(14)$ \\
$[1.1,6]$ & $0.0857(78)$ & $-0.084(84)$ & $0.19(21)$   & $0.036(42)$   & $0.004(15)$ \\
$[15,16]$ & $0.0231(33)$ & $-0.012(12)$ & $-0.022(40)$ & $-0.070(15)$  & $-0.0105(40)$ \\
$[16,18]$ & $0.0171(24)$ & $-0.010(12)$ & $-0.011(24)$ & $-0.106(12)$  & $-0.0104(30)$ \\
$[18,20]$ & $0.0094(16)$ & $-0.005(10)$ & $-0.003(9)$  & $-0.1713(80)$ & $-0.0086(20)$ \\
$[15,20]$ & $0.0157(22)$ & $-0.009(11)$ & $-0.010(22)$ & $-0.121(11)$  & $-0.0098(27)$ \\
\hline \hline
\end{tabular}
\end{center}
\end{table}
%%%%%%%%%%%%%%%%%%%%%%%%%%%%%%%%%%%%%%%%%%%%%%%%%%%%%%%

As can be seen from Figure~\ref{fig:a2hdm} and Table~\ref{tab:binSM}, in the bin $[15, 20]~{\rm GeV^2}$ where both the experimental data and the lattice QCD predictions for the $\Lambda_b\to \Lambda$ transition form factors are most precise, the measured differential branching fraction~\cite{Aaij:2015xza}, $(1.20\pm0.27)\times10^{-7}~{\rm GeV}^{-2}$, exceeds the SM prediction, $(0.766\pm0.069)\times10^{-7}~{\rm GeV}^{-2}$, by about $1.6\,\sigma$. Although being not yet statistically significant, it is interesting to note that the deviation is in the opposite direction to what has been observed in the $B\to K^{(\ast)}\mu^+\mu^-$~\cite{Aaij:2013iag,Aaij:2014pli,Bouchard:2013mia} and $B_s\to\phi\mu^+\mu^-$~\cite{Aaij:2013aln,Aaij:2015esa,Horgan:2013pva} decays, where the measured differential branching fractions favor, on the other hand, smaller values than their respective SM predictions. Also in this bin, the lepton-side forward-backward asymmetry measured by LHCb~\cite{Aaij:2015xza}, $-0.05\pm0.09$, is found to be about $3.3\,\sigma$ higher than the SM value, $-0.349\pm0.013$. As detailed in Ref.~\cite{Meinel:2016grj}, combing the current data for $\Lambda_b\to \Lambda(\to p\pi^-)\mu^+\mu^-$ decay with that for the branching ratios of $B_s\to\mu^+\mu^-$ and inclusive $b\to s \ell^+\ell^-$ decays, Meinel and Dyk found that their fits prefer a positive shift to the Wilson coefficient $C_9$, which is opposite in sign compared to that found in the latest global fits of only mesonic decays~\cite{Altmannshofer:2014rta,Descotes-Genon:2015uva,Hurth:2016fbr}. This means that a simple shift in $C_9$ alone could not explain all the current data. Especially, a negative shift in $C_9$, as found in global fits of only mesonic observables, would further lower the predicted $\Lambda_b\to\Lambda \mu^+\mu^-$ differential branching fraction.

Our SM predictions for the zero-crossing points of angular observables $A_{\rm FB}^\ell$, $A_{\rm FB}^{\ell\Lambda}$ and $Y_2$ read, respectively, as
\begin{align}\label{eq:q0-SM}
 q_0^2(A_{\rm FB}^\ell)|_{\rm SM} &=(3.95\pm0.62)~{\rm GeV}^2\,, \qquad
 q_0^2(A_{\rm FB}^{\ell\Lambda})|_{\rm SM}=(3.89\pm0.63)~{\rm GeV}^2\,, \nonumber\\[0.2cm]
 q_0^2(Y_2)|_{\rm SM} &=(0.35\pm0.10)~{\rm GeV}^2\,.
\end{align}
The zero-crossing points of the other observables $Y_i$~($i={\rm 3s,\,3sc,\,4s,\,4sc})$, which correspond to the case when the relative angular momentum between the $p\pi^-$ system and the dimuon system is $(l, m)=(1,\pm1)$, are plagued by large theoretical uncertainties. The observables $Y_{\rm 3s}$ and $Y_{\rm 3sc}$ are predicted to be very small within SM and are, therefore, potentially good probes of NP beyond the SM~\cite{Kumar:2015tnz}.

\subsection{Results in the A2HDM}

In this subsection, we shall investigate the impact of A2HDM on the $\Lambda_b\to\Lambda(\to p\pi^-) \mu^+\mu^-$ observables. For simplicity, the alignment parameters $\varsigma_{u,d}$ are assumed to be real. As in our previous paper~\cite{Hu:2016gpe}, we use the inclusive $B\to X_s\gamma$ branching fraction~\cite{Amhis:2014hma,Misiak:2015xwa} and the latest global fit results of $b\to s\ell\ell$ data~\cite{Descotes-Genon:2015uva,Meinel:2016grj} to restrict the model parameters $\varsigma_{u,d}$. Under these constraints, numerically, the charged-scalar contributions to the Wilson coefficients can be divided into the following two cases~\cite{Hu:2016gpe}:
\begin{align*}
&\mbox{$\mbox{Case A}$:\quad $C_{7,9,10}^\mathrm{H^\pm}$ are sizable, but $C_{9,10}^{\prime\mathrm{H^\pm}}\simeq0$}\,;\\[0.3cm]
&\mbox{$\mbox{Case B}$:\quad $C_7^\mathrm{H^\pm}$ and $C_{9,10}^{\prime\mathrm{H^\pm}}$ are sizable, but $C_{9,10}^\mathrm{H^\pm}\simeq0$}\,.
\end{align*}
They are associated to the (large $\left|\varsigma_u\right|$, small $\left|\varsigma_d\right|$) and (small $\left|\varsigma_u\right|$, large $\left|\varsigma_d\right|$) regions, respectively; see Ref.~\cite{Hu:2016gpe} for more details. This means that the charged-scalar exchanges contribute mainly to left- and right-handed semileptonic operators in case~A and case~B, respectively. The influences of these two cases on the $\Lambda_b\to\Lambda(\to p\pi^-) \mu^+\mu^-$ observables are shown in Figure~\ref{fig:a2hdm}, where the blue (in case~A) and red (in case~B) bands are obtained by varying randomly the model parameters within the ranges allowed by the global fits~\cite{Descotes-Genon:2015uva,Meinel:2016grj,Hu:2016gpe}, with all the other input parameters taken at their respective central values.

%%%%%%%%%%%%%%%%%%%%%%%%%%%%%%%%%%%%%%%%%%%%%%%%%%%%%%%
\begin{table}[t]
\begin{center}
\caption{\label{tab:sma2hdmbin}Comparison of our results for the $\Lambda_b\to\Lambda(\to p\pi^-) \mu^+\mu^-$ observables with the LHCb data (where available) in the bin $[15,20]~{\rm GeV}^2$. The uncertainties of A2HDM results mainly come from the $\Lambda_b\to\Lambda$ transition form factors and the model parameters. The differential branching fraction is given in units of $10^{-7}~{\rm GeV}^{-2}$.}
\vspace*{0.3cm} \tabcolsep 0.12in
\renewcommand{\arraystretch}{1.0}
\begin{tabular}{cccccc}
\hline \hline
& $\langle \mathrm{d}\mathcal{B}/\mathrm{d} q^2\rangle$
& $\langle F_L \rangle $
& $\langle A_{\rm FB}^\ell \rangle $
& $\langle A_{\rm FB}^\Lambda \rangle $
& $\langle A_{\rm FB}^{\ell\Lambda} \rangle $ \\
\hline
SM             & $0.766(69)$ & $0.409(13)$
& $-0.349(13)$ & $-0.2709(73)$ & $0.1394(32)$ \\
Case~A   & $0.82(11)$ & $0.409(13)$
& $-0.344(16)$ & $-0.2709(73)$ & $0.1374(47)$ \\
Case~B   & $1.11(38)$ & $0.396(20)$
& $-0.24(12)$ & $-0.179(92)$ & $0.129(17)$ \\
LHCb~\cite{Aaij:2015xza}   & $1.20(27)$ & $0.61^{+0.11}_{-0.14}$
& $-0.05(9)$ & $-0.29(8)$ & --- \\
\hline \hline
& $\langle Y_2 \rangle $
& $\langle Y_{\rm3\,s} \rangle\times10^{-2} $
& $\langle Y_{\rm3\,sc} \rangle\times10^{-3} $
& $\langle Y_{\rm4\,s} \rangle $
& $\langle Y_{\rm4\,sc} \rangle $ \\
\hline
SM           & $0.0157(22)$ & $-0.009(11)$
& $-0.010(22)$ & $-0.121(11)$  & $-0.0098(27)$ \\
Case~A & $0.0156(23)$ & $-0.008(11)$
& $-0.011(23)$ & $-0.120(11)$  & $-0.0097(27)$ \\
Case~B & $0.0110(56)$ & $-0.68(68)$
& $-0.035(39)$ & $-0.174(55)$  & $-0.007(4)$ \\
\hline \hline
\end{tabular}
\end{center}
\end{table}
%%%%%%%%%%%%%%%%%%%%%%%%%%%%%%%%%%%%%%%%%%%%%%%%%%%%%%%

In case~A, the impact of A2HDM is found to be negligibly small on the hadron-side forward-backward asymmetry $A_{\rm FB}^\Lambda$ and the observables $Y_i$~($i={\rm 3s,\,3sc,\,4s,\,4sc})$. For the differential branching fraction, on the other hand, visible enhancements are observed relative to the SM prediction, especially in the high $q^2$ region. For the remaining observables, the A2HDM only affects them in the low $q^2$ region, but the effect is diluted by the SM uncertainty. In order to see clearly the A2HDM effect in case~A, we give in Table~\ref{tab:sma2hdmbin} the values of the binned observables in the bin $[15,20]~{\rm GeV}^2$, including also the SM predictions, the A2HDM effect in case~B, as well as the LHCb data (where available) for comparison. Although being improved a little bit, the deviations between the LHCb data and the theoretical values for the differential branching fraction and the lepton-side forward-backward asymmetry are still at $1.3\,\sigma$ and $3.2\,\sigma$, respectively. Including the A2HDM in case~A, there are only small changes on the zero-crossing points:
\begin{align}\label{eq:q0-case-A}
q_0^2(A_{\rm FB}^\ell)|_{\rm case~A} &= (4.02\pm1.01)\,{\rm GeV}^2\,, \qquad
q_0^2(A_{\rm FB}^{\ell\Lambda})|_{\rm case~A}=(3.96\pm1.02)\,{\rm GeV}^2\,, \nonumber\\[0.2cm]
q_0^2(Y_2)|_{\rm case~A} &= (0.37\pm0.20)\,{\rm GeV}^2.
\end{align}

In case~B, however, the A2HDM has a significant influence on almost all the observables, as shown in Figure~\ref{fig:a2hdm}. The most prominent observation is that it can enhance both the differential branching fraction and the lepton-side forward-backward asymmetry in the bin $[15,20]~{\rm GeV}^2$, being now compatible with the experimental measurements at $0.2\,\sigma$ and $1.3\,\sigma$, respectively~(see also Table~\ref{tab:sma2hdmbin}). The magnitude of the hadron-side forward-backward asymmetry tends to become smaller in the whole $q^2$ region in this case, but is still in agreement with the LHCb data, with the large experimental and theoretical uncertainties taken into account. In the high~(whole) $q^2$ region, a large effect is also observed on the asymmetry observable $Y_{\rm 3s}$~($Y_{\rm 4s}$). Adding up the A2HDM effect in case~B, the zero-crossing points are now changed to
\begin{align}\label{eq:q0-case-B}
q_0^2(A_{\rm FB}^\ell)|_{\rm case~B} &=(4.38\pm1.44)\,{\rm GeV}^2\,, \qquad q_0^2(A_{\rm FB}^{\ell\Lambda})|_{\rm case~B} =(4.00\pm1.17)\,{\rm GeV}^2\,,\nonumber\\[0.2cm]
q_0^2(Y_2)|_{\rm case~B} &=(0.52\pm0.29)\,{\rm GeV}^2\,,
\end{align}
which are all significantly enhanced compared to the SM predictions (see Eq.~\eqref{eq:q0-SM}) and the results in case~A (see Eq.~\eqref{eq:q0-case-A}). It should be noticed that our predictions for the zero-crossing points given by Eqs.~\eqref{eq:q0-SM}--\eqref{eq:q0-case-B} are most severely affected by the hitherto unquantified systematic uncertainty coming from the non-factorizable spectator-scattering contributions at large hadronic recoil, a caveat emphasized already in sec.~\ref{subsec:Heff}.

Combining the above observations with our previous studies---the angular observable $P_5'$ in $B^0\to K^{\ast 0}\mu^+\mu^-$ decay could be increased significantly to be consistent with the experimental data in case~B~\cite{Hu:2016gpe}, we could, therefore, conclude that the A2HDM in case~B is a promising alternative to the observed anomalies in $b$-hadron decays.

\section{Conclusions}
\label{sec:conclusion}

In this paper, we have investigated the A2HDM effect on the rare baryonic  $\Lambda_b\to\Lambda(\to p\pi^-)\mu^+\mu^-$ decay, which is mediated by the same quark-level $b\to s\mu^+\mu^-$ transition as in the mesonic $B\to K^{(\ast)}\mu^+\mu^-$ decays. In order to extract all the ten angular coefficients, we have considered the differential branching fraction ${\rm d}{\cal B}/{\rm d}q^2$, the longitudinal polarization fraction $F_L$, the lepton-, hadron- and combined lepton-hadron-side forward-backward asymmetries $A_{\rm FB}^\ell$, $A_{\rm FB}^\Lambda$ and $A_{\rm FB}^{\ell\Lambda}$, as well as the other five asymmetry observables $Y_i$~($i={\rm 2,\,3s,\,3sc,\,4s,\,4sc}$). For the $\Lambda_b\to\Lambda$ transition form factors, we used the most recent high-precision lattice QCD calculations with $2+1$ dynamical flavours.

Taking into account constraints on the model parameters $\varsigma_{u,d}$ from the inclusive $B\to X_s\gamma$ branching fraction and the latest global fit results of $b\to s\ell\ell$ data, we found numerically that the charged-scalar exchanges contribute either mainly to the left- or to the right-handed semileptonic operators, labelled by case~A and case~B, respectively. The influences of these two cases on the $\Lambda_b\to\Lambda(\to p\pi^-) \mu^+\mu^-$ observables are then investigated in detail. While there are no significant differences between the SM predictions and the results in case~A, the A2HDM in case~B is much favored by the current data. Especially in the bin $[15, 20]~{\rm GeV^2}$ where both the experimental data and the lattice QCD predictions are most precise, the deviations between the SM predictions and the experimental data for the differential branching fraction and the lepton-side forward-backward asymmetry could be reconciled to a large extend. Also in our previous paper~\cite{Hu:2016gpe}, we have found that the angular observable $P_5'$ in $B^0\to K^{\ast 0}\mu^+\mu^-$ decay could be increased significantly to be consistent with the experimental data in case~B. Therefore, we conclude that the A2HDM in case~B is a very promising solution to the currently observed anomalies in $b$-hadron decays.

Finally, it should be pointed out that more precise experimental measurements of the full angular observables, especially with a finer binning, as well as a systematic analysis of non-factorizable spectator-scattering effects in $\Lambda_b\to\Lambda(\to p\pi^-)\mu^+\mu^-$ decay, would be very helpful to further deepen our understanding of the quark-level $b\to s\mu^+\mu^-$ transition.

\section*{Acknowledgements}

The work is supported by the National Natural Science Foundation of China (NSFC) under contract Nos.~11675061, 11435003, 11225523 and 11521064. QH is supported by the Excellent Doctorial Dissertation Cultivation Grant from CCNU, under contract number 2013YBZD19.

\appendix

\section{Wilson coefficients in A2HDM}
\label{app:WCA2HDM}

The coefficients $C_{i,{\rm XY}}^{(\prime)}$~($i=7,9,10$ and ${\rm X,Y=u,d}$) appearing in the Wilson coefficients $C_{7,9,10}^{(\prime)\mathrm{H^\pm}}$ are given, respectively, as~\cite{Hu:2016gpe}
\begin{align} \label{eq:WCA2HDM1}
&C_{7,\,\mathrm{uu}}=-\frac{1}{6}F_{6}(y_{t})\,,\\[0.2cm]  \label{eq:WCA2HDM2} &C_{7,\,\mathrm{ud}}=-\frac{4}{3}F_{1}(y_{t})-\frac{80}{17}F_{2}(y_{t}) -\frac{3}{17}F_{5}(y_{t})+\frac{1}{17}F_{6}(y_{t})\,,\\[0.2cm] \label{eq:WCA2HDM3}
&C_{9,\,\mathrm{uu}}=\frac{8}{9}F_{1}(y_{t})-\frac{896}{51}F_{2}(y_{t}) -\frac{1}{17}F_{5}(y_{t})-\frac{14}{153}F_{6}(y_{t}) -\frac{x_{t}}{2}\left(-4+\frac{1}{\sin^{2}\theta_{W}}\right)F_{1}(y_{t})\,,\\[0.2cm] \label{eq:WCA2HDM4}
&C_{10,\,\mathrm{uu}}=\frac{x_{t}}{2\sin^{2}\theta_{W}}F_{1}(y_{t})\,,\\[0.2cm] \label{eq:WCA2HDM5}
&C_{9,\,\mathrm{uu}}'=\frac{y_{t}}{x_{t}}F_{8}(y_{t})\,,\\[0.2cm] \label{eq:WCA2HDM6}
&C_{9,\,\mathrm{ud}}'=\frac{y_{t}}{x_{t}}F_{7}(y_{t})\,,\\[0.2cm] \label{eq:WCA2HDM7}
&C_{9,\,\mathrm{dd}}'=\frac{y_{t}}{x_{t}}\left[\frac{2}{9}F_{0}\left(x_{t}\right) +\frac{20}{9}F_{1}(y_{t})+\frac{928}{51}F_{2}(y_{t}) -\frac{2}{17}F_{5}(y_{t})-\frac{11}{153}F_{6}(y_{t})\right]\,,\\[0.2cm] \label{eq:WCA2HDM8}
&C_{10,\,\mathrm{uu}}'=-\frac{1}{17}\left(80F_{2}(y_{t})+3F_{5}(y_{t}) -F_{6}(y_{t})\right)\,,\\[0.2cm] \label{eq:WCA2HDM9}
&C_{10,\,\mathrm{ud}}'=\frac{1}{\sin^{2}\theta_{W}}\left[-\frac{1}{12}F_{1}(y_{t}) +\frac{30}{17}F_{2}(y_{t})+\frac{9}{136}F_{5}(y_{t}) -\frac{3}{136}F_{6}(y_{t})\right]\, \nonumber \\[0.1cm]
& \hspace{1.6cm} -\frac{1}{6}\left(-4+\frac{1}{\sin^{2}\theta_{W}}\right)F_{1} (y_{t})\,,\\[0.2cm] \label{eq:WCA2HDM10}
&C_{10,\,\mathrm{dd}}'=-\frac{1}{\sin^{2}\theta_{W}}\left[\frac{1}{2}F_{1}(y_{t}) +F_{2}(y_{t})\right]+\left(-4+\frac{1}{\sin^{2}\theta_{W}}\right)F_{2}(y_{t})\,,
\end{align}
where the basic functions $F_i(x)$ are defined, respectively, by
\begin{align} \label{eq:BF1}
F_{0}(x)&=\ln x\,,\\[0.2cm] \label{eq:BF2}
F_{1}(x)&=\frac{x}{4-4x}+\frac{x\ln x}{4(x-1)^{2}}\,,\\[0.2cm] \label{eq:BF3}
F_{2}(x)&=\frac{x}{96(x-1)}-\frac{x^{2}\ln x}{96(x-1)^{2}}\,,\\[0.2cm] \label{eq:BF4}
F_{3}(x)&=\frac{x}{8}\left[\frac{x-6}{x-1}+\frac{(3x+2)\ln x}{(x-1)^{2}}\right]\,,\\[0.2cm] \label{eq:BF5}
F_{4}(x)&=-\frac{3x(x-3)}{32(x-1)}+\frac{x\left(x^{2}-8x+4\right)\ln x}{16(x-1)^{2}}\,,\\[0.2cm] \label{eq:BF6}
F_{5}(x)&=\frac{-19x^{3}+25x^{2}}{36(x-1)^{3}}+\frac{\left(5x^{2}-2x-6\right)x^{2}\ln x}{18(x-1)^{4}}\,,\\[0.2cm] \label{eq:BF7}
F_{6}(x)&=\frac{8x^{3}+5x^{2}-7x}{12(x-1)^{3}}-\frac{(3x-2)x^{2}\ln x}{2(x-1)^{4}}\,,\\[0.2cm] \label{eq:BF8}
F_{7}(x)&=\frac{x\left(53x^{2}+8x-37\right)}{108(x-1)^{4}}+\frac{x\left(-3x^{3}-9x^{2}+6x+2\right)\ln x}{18(x-1)^{5}}\,,\\[0.2cm] \label{eq:BF9}
F_{8}(x)&=\frac{x\left(18x^{4}+253x^{3}-767x^{2}+853x-417\right)}{540(x-1)^{5}}-\frac{x\left(3x^{4}-6x^{3}+3x^{2}+2x-3\right)\ln x}{9(x-1)^{6}}\,.
\end{align}

\bibliographystyle{JHEP}
\bibliography{references}

\end{document}